\documentclass{llncs}
\usepackage{makeidx}  
\usepackage{graphicx}
\usepackage{amsmath,amssymb} 
\newcommand{\ket}[1]{ \left |#1\right \rangle}
\makeindex

\begin{document}

\frontmatter          
\setcounter{page}{1}

\pagestyle{headings}  

\title{Complexity of Graph State Preparation}
\author{Mehdi Mhalla, Simon Perdrix}
\institute{Leibniz Laboratory\\ 46, avenue F\'elix Viallet 38000 Grenoble, France\\ \emph{mehdi.mhalla@imag.fr,simon.perdrix@imag.fr}}

\maketitle

\begin{abstract}
The graph state formalism is a useful abstraction of entanglement. It is used in some multipartite purification schemes  and it adequately represents universal resources for measurement-only quantum computation.
We focus in this paper on the complexity of graph state preparation.  We consider the number of ancillary qubits, the size of the primitive operators, and the duration of preparation. For each lexicographic order over these parameters we  give upper and lower bounds for the complexity of graph state preparation.
The first part motivates our work and introduces  basic notions and notations for the study of graph states.
Then we study some graph properties of graph states, characterizing their minimal degree by local unitary transformations, we propose an algorithm to reduce the degree of a graph state, and  show the relationship with Sutner $\sigma$-game.
  These properties are used in the  last part, where  algorithms and lower bounds for each lexicographic order over the considered parameters are presented.
\end{abstract}

\section{Graph States}

For any graph $G=(V,E)$, a graph state $\ket{G}$ is the unique state (up to a global phase) such that $\forall v\in V$, $g_v\ket{G}=\ket{G}$, with  $g_v$  the Pauli operator $X^{(v)}\prod_{k\in N_G(v)} Z^{(k)}$ where $N_G(v)$ is the set of vertices adjacent to $v$.

The set of operators $\{ g_v,v\in V\}$, which form a group, is called a \emph{stabilizer} \cite{NC00} of $\ket{G}$.

 For example, the graph :
\begin{center}
\includegraphics[width=1cm]{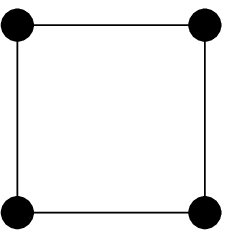}
\end{center}

corresponds to the state characterized by a stabiliser that can be represented as follows: 
$$
\left(
\begin{array}{cccc}
X&Z&I&Z \\
Z&X&Z&I \\
I&Z&X&Z \\
Z&I&Z&X
\end{array}
\right)
$$

where each column corresponds to a qubit, and each row is a Pauli operator. 

Since the action of local unitary transformations does not modify, \emph{a priori} the entanglement structure of a quantum state, a relationship over quantum states, called \emph{LU-equivalence} can be stated as follows: two quantum states are \emph{LU-equivalent} iff these states are equal up to a local unitary transformation. In a same way two quantum states are \emph{LP-equivalent} iff these states are equal up to a Pauli transformation.

Even if for a given graph $G$ with $n$ vertices, there exist $2^n$ different quantum states which are LP-equivalent to $\ket{G}$, only one of them is a graph state: $\forall G,G', \ket{G}\equiv_{LP}\ket{G'} \iff G=G'$. This property cannot be generalized to the LU-equivalent case because there exist some graphs $G$, $G'$ such that $G\neq G'$ and $\ket{G}\equiv_{LU}\ket{G'}$. 

The LU-equivalent class of a given graph $G$ can be characterized by Theorem \ref{vdn} which is due to  Van den Nest \cite{VDM03}. This characterization is based on a  classical graph operation called local complementation. 

\begin{figure}
\begin{center}
\includegraphics[width=0.8\textwidth]{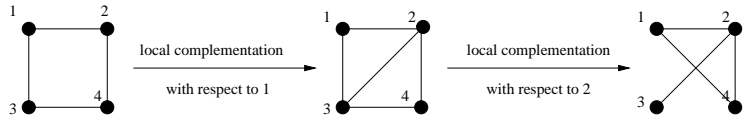}
\label{c4}
\caption {Local complementation from graph $C_4$}
\end{center}
\end{figure}

This operation, introduced by Bouchet \cite{B88,B91}, was originally used to characterize circle graphs \cite{B94}. It can be stated as follows: the local complementation with respect to a vertex $v$ in  some graph $G=(V,E)$ is the graph $G'=(V,E\Delta K_v )$ where $\Delta$ is the symmetric difference operation and $K_v$ is the complete graph over $N_G(v)$ (the neighborhood of $v$ in $G$), so the neighborhood of $v$ is complemented in $G$. An example of local complementation is presented in Figure \ref{c4}.
 
\begin{theorem}[Van den Nest] 
\label{vdn}
Given two graphs $G$ and $G'$, $\ket{G}$ and $\ket{G'}$ are LU-equivalent iff there exists a sequence of local complementations which transforms $G$ into $G'$.
\end{theorem}

Equivalence of graph states has also been studied by Eisert and Schlingemann \cite{HEB03,S03}.   Another relevant subject concerning graph states is the consumption of graph states, \emph{i.e.} the application of one-qubit measurements on a given graph state. A computational model based on the consumption of graph states, called one-way quantum computation, has been proved universal for quantum computation \cite{R03} . The purpose of this paper is a dual approach to graph state consumption, which is graph state preparation: for a given graph $G$, the preparation of $\ket{G}$ consists in applying some operations, like unitary transformations or projective measurements, in order to obtain $\ket{G}$ starting from an initial state, which is an unknown state if the projective measurement are allowed and a known but not entangled state otherwise.  Leung \cite{L04} has proposed an algorithm which permit, for any graph $G$, to prepare the corresponding graph state $\ket{G}$ using projective measurements only. The relevant parameters which permits to measure the complexity of a graph state preparation are: the number of ancillary qubits, the size of the primitive operators (unitary transformations or projective measurements), and the duration of the preparation. For each lexicographic order over these parameters, we give an upper and a lower bound of the complexity of graph state preparation.

\section{Local Complementation and Minimal Degree}
\label{grph}
In this section, the relationship between local complementation and the minimal degree $\delta(G)$ of a graph $G$, \emph{i.e.} the minimal vertex degree in $G$, is explored. The results in this section will be used to characterize the complexity of some graph state preparations which depend on the \emph{minimal degree under local complementation} of the graph to prepare.

For a given  graph $G=(V,E)$:
\begin{itemize}
\item $G'=\lambda_v(G)$ is the graph obtained from $G$ by a local complementation with respect to the vertex $v \in V$.
\item the {\bf minimal degree under local complementation} of $G$, $\delta_{loc}(G)$, is the smallest minimal degree of a graph that can be reached by a sequence of local complementations from $G$.
\item  a subsest $D\subseteq V$ is an {\bf evenly seen set} if $\forall u\in V\setminus D, |N_G(u) \bigcap D|$ is even.
\item  a subsest $K\subseteq V$ is called {\bf $d$-locally evenly seen} if there exists a set $D$ with $K \subseteq D \subseteq V$ and $|D|=d$, such that $\forall u \in V\setminus D, |N_G(u) \bigcap K|$ is even.  
\end{itemize}

The following lemmas exhibit upper bounds for the minimal degree under local complementation.

\begin{lemma}
If a graph $G$ has an evenly seen set $D$ such that $|D|=d$, then $\delta_{loc}(G) \le d-1$
\end{lemma}

This lemma is a special instance of the following more general one:

\begin{lemma}
\label{thm:car}
If a graph $G$ has a $d$-locally evenly seen set then  $\delta_{loc}(G) \le d-1$.
\end{lemma}
\begin{proof}

Let $G=(V,E)$ be a graph, $D$ be a subset of $V$ and $K$ be a subset of $D$ with $|D|=d$ and $|K|=k$, such that $\forall u \in V\setminus D, |N_G(u) \bigcap K|$ is even.

The proof is by induction on $k=|K|$. 

Suppose that the set $K$ contains only one vertex, $i.e.$ $K=\{v\}$. Since $\forall u \in V\setminus D,|N_G(u) \bigcap \{v\}|$ is even, all the neighbors of $v$ are in $D$ and thus $\delta_{loc}(G) \le d-1$.

We will now prove that if a graph $G=(V,E)$ has  a $d$-locally evenly seen set $K$, with 
$k=|K|\ge 2$, then either $\delta(G) \le d-1$ or a graph $G'$, which has a $d$-locally evenly seen set $K'$ with $|K'|=k-1$, can be reached from $G$ by means of local complementations.

  Suppose that there exists a vertex $v$ in $K$ such that the cardinality of $N_0=N_G(v) \bigcap K$ is odd, we claim that the graph $G'=\lambda_v(G)$ is such that $\forall u \in V\setminus D,|N_{G'}(u) \bigcap K'|$ is even, where $K'=K\setminus \{v\}$ (Figure 2).
  
  Indeed, for any vertex  $u \in  V \setminus D$:
\begin{itemize}
\item if $u \notin N_G(v)$ then
 $|N_{G'}(u) \bigcap K'|=|N_{G'}(u) \bigcap K|=|N_G(u) \bigcap K|=0 \mod 2$, 
\item  if $u \in N_G(v) $, then 
$N_G(u) \bigcap K=\{v\} \bigcup N_1 \bigcup N_2$, where $N_1=N_G(u) \bigcap N_0$ is the  set  of common neighbors of $u$ and $v$ in $K$ and $N_2=N_G(u) \bigcap (K \setminus (N_G(v) \bigcup \{v\}))$ is the set of neighbors of $u$ different from $v$, that are not neighbors of $v$. By hypothesis, $1+ |N_1| +|N_2|$ is even. After the local complementation, $N_{G'}(u)\bigcap K'=(N_0 \setminus N_1) \bigcup N_2$. Since $N_0$ is odd, $(N_0 \setminus N_1) $ has a parity opposite to $|N_1|$, and $| N_{G'}(u)\bigcap K'|$ is even.
   \end{itemize}
   
   Thus, by induction,  $\delta_{loc}(G) \le d-1$.
   \begin{figure}
   \begin{center}
\includegraphics[width=8cm]{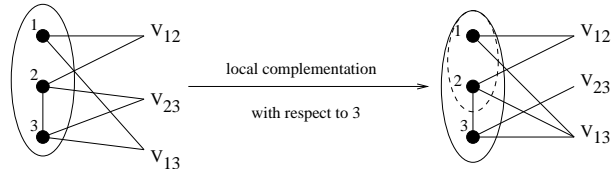}

\caption{Local complementation with respect to a vertex with odd degree in $K$. $v_{ij}$ represents the set of common neighbors to vertices $i$ and $j$.}
\label{fg1}
\end{center}
\end{figure}

   Now suppose that   all vertices $v$ in $K$ are such that $| N_G(v) \bigcap K| $ is even. 
   
   We show that a local complementation with respect to a vertex in $V\setminus D$ that has a neighbor $v\in K$  changes the parity of $|N_G(v) \bigcap K|$ (Figure \ref{fg2}). 
      
   Consider a vertex $v$ in $K$ and let $N_0=N_G(v) \bigcap K$ be its neighborhood inside $K$.
   \begin{itemize}
   \item If $v$  has no neighbor outside $D$ then its degree is at most $d-1$ and  $\delta_{loc}(G) \le d-1$.
   \item Otherwise, there exists a vertex $u$ in $V \setminus D$ such that $v\in N_G(u) \bigcap K$ and $|N_G(u) \bigcap K|$ is even. Then, $N_G(u) \bigcap K= \{v\} \bigcup N_1 \bigcup N_2$ where $N_1=N_G(u) \bigcap N_0$ is the set  of common neighbors of $u$ and $v$ in $K$, and $N_2=N_G(u) \bigcap (K \setminus (N_G(v) \bigcup \{v\}))$ is the set of neighbors of $u$ that are not neighbors of $v$. By hypothesis, $1+ |N_1| +|N_2|$ is even. With $G'=\lambda_u(G)$, $N_{G'}(v) \bigcap K = ((N_G(v) \bigcap K) \setminus N_1) \bigcup N_2$ is odd. Besides,  $K$ is still evenly  seen  by the vertices outside $D$ in $G'$. Indeed, for any vertex $u_1 \in V\setminus D$,  $N_{G'}(u_1)\bigcap K=
  (N_{G}(u_1)\bigcap K) \Delta (N_G(u) \bigcap K )$. $\hfill \Box$
\end{itemize}
\begin{figure}
   \begin{center}
\includegraphics[width=8cm]{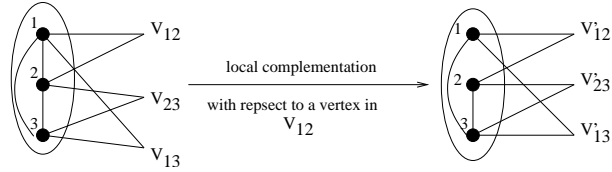}

\caption{  Local complementation with respect to a vertex outside $D$}
\label{fg2}
\end{center}
\end{figure}
\end{proof}

Note that the previous proof is a constructive one, and it defines an algorithm which, given a graph $G$ and a $d$-locally evenly seen set $K$, outputs a sequence of at most $2|K|$ local complementations $S=\{\lambda_{v_1} \ldots \lambda_{v_s}\}$, such that $\delta(\lambda_{v_s} \circ \ldots \circ \lambda_{v_1}(G))\le d-1$.
\newpage
The algorithm for reducing degree is the following: 

\begin{itemize}
\item Initially S=\{\} is the sequence of local complementations.
\item While $\delta(G)> d-1$:
  \subitem if there exists a vertex $v\in K$, such that  $|N_G(v)\bigcap K|$ is  odd, then  $S=S \bigcup \{\lambda_v\}$, $G=\lambda_v(G)$, $K= K-\{v\}$,
    \subitem   else   $ S=S \bigcup \{\lambda_u\}, G=\lambda_u(G)$, where $u$ is a vertex in $G\setminus K$ such that $|N_G(u) \bigcap K|$ is even.
\end{itemize}
   
 The following claim contributes to the proof of the upper bound:
\begin{claim}
\label{cl1}
If $\delta(G) \le d$ then $G$ has a $(d+1)$-locally evenly seen set.
\end{claim}
\begin{proof}
Let $v$ be a vertex of degree $d$, $K=\{v\}$ and $D=\{v\}\bigcup N_G(v)$. $K$ is a $(d+1)$-locally evenly seen set. $\hfill \Box$
\end{proof}

\begin{lemma}
If a graph $G$ has no $d$-locally even seen set then $\delta_{loc}(G) \ge d$.
\end{lemma}
\begin{proof}
Using  the previous claim, it is sufficient  to prove that  a $d$-locally evenly seen set cannot be created by means of  local complementations.

By contradiction, suppose that $G$ has no $d$-locally evenly seen set and $G'=\lambda_v(G)$ has a  $d$-locally evenly seen set $K$.

Notice that a local complementation is its self inverse: $\lambda_v\lambda_v(G)=G$. 

\begin{itemize}
\item If $v\in V\setminus D$  then $K$ is still a  $d$-locally evenly seen set in $\lambda_v(G')=G$  (as in proof of lemma \ref{thm:car}) which is impossible.
\item If $v\in D$, let $L=N_G(v) \bigcap K$: 
\subitem if $|L|$ is even, for any vertex $u\in V\setminus D$ the parity of $|N_G(u)\bigcap K|$ does not change by local complementation with respect to $v$ which is impossible.
\subitem if $|L|$ is odd then, if $v\in K$ then $K \setminus \{v\}$ is $d$-locally evenly seen by complementing with respect to  $v$ (as in proof of lemma \ref{thm:car}). Otherwise, $v$ has an odd number of neighbors in $V  \setminus D$ and all its neighbors evenly see $K$. Thus after the local complementation $K\bigcup \{v\}$ is a $d$-locally evenly seen set in $\lambda_v(G')$ which is impossible. $\hfill \Box$
\end{itemize} 
\end{proof}
\begin{theorem}
\label{thf}
A graph $G$ has a $d$-locally even seen set and no $(d-1)$-locally even seen set \emph{iff}
 $\delta_{loc}(G)=d-1$. 
 \end{theorem}
\begin{proof}
Just combine the two previous lemmas. $\hfill \Box$
\end{proof}
Examples :
\begin{itemize}
\item  The graph $C_4$ (Figure \ref{c4}) has a $2$-locally evenly seen set ($K=D=\{1,3\}$) and no $1$-locally evenly seen set, so Theorem \ref{thf} implies that $\delta_{loc}(C_4)=1$.
\item An $n\times m$ cluster state (grid graph) with $n\ge 2$ or $m\ge2$ has as minimal degree $2$ under local complementation.
\end{itemize}

Theorem \ref{thf} can be formulated as an optimisation problem:
\begin{corollary}
For a given graph $G=(V,E)$,
$\delta_{loc}(G)=|V|-1-\max_{K\subseteq V, K\neq \emptyset}|\{ v \in V\setminus K , |N_G(v)\bigcap K| \equiv 0 \mod 2 \}|$.
\end{corollary}

Furthermore, thanks to the  degree reducing algorithm, finding an optimal set $K$ permits to derive an optimal sequence of local complementations to reduce the  degree of a graph $G$ to $\delta_{loc}(G)$.
However, to our knowledge, the complexity of this optimisation  is still an open problem.

Note that Lemma \ref{thm:car} can be seen in terms of a combinatorial game on graphs called \emph{$\sigma$-game}. This game was  introduced by Sutner \cite{S89} generalised in \cite{GMT03} and is related to cellular automata. A configuration on a graph is an assignment of values in
$\{0,1\}$ 
to the vertices in $G$. One may think of a vertex $v$ of $G$ as a
button the player can press at his discretion. If vertex $v$ is
chosen, the value of all the vertices adjacent to $v$ are flipped.
 
Indeed,  a set $K$ is $d$-locally seen \emph{iff} playing on all the  vertices in $K$ does not modify the state of the vertices in $V\setminus D$. It comes:   

\begin{theorem}
\label{thf2}
For a given graph $G$ such that $\delta(G)=\delta_{loc}(G)=d$ and for any set $D$ of $d$ vertices, all the possible configurations on $D$ can be reached by playing only on vertices outside $D$.  
\end{theorem}

\begin{proof}
Since  $\delta(G)=\delta_{loc}(G)$, any subset $D$ of $d$ vertices has no subset $K$ such that pressing on the vertices in $K$ does not modify $D\setminus K$. 

Consider  the $(n-d) \times d$ adjacency  matrix $A_D$, where the lines correspond to the vertices in $V\setminus D$, and the columns to the vertices in $D$.
If $X\in {\mathbb F}_2^d$ is a vector representing the pressed vertices in $D$, $A_D .X$ is the characteristic vector of the flipped vertices in $V\setminus D$.
\begin{figure}
$$
\begin{array}{cl}
& \begin{array}{ccc} \, \,i_1&i_2&i_3 \end{array} \\
\begin{array}{c}
e_1\\
e_2\\
e_3\\
e_4\\
\end{array} &
\left[
\begin{array}{ccc}
1&0&1\\
0&1&1\\
0&0&0\\
1&1&0\\
\end{array}
\right]
.
\left[
\begin{array}{c}
0\\
1\\
1
\end{array}
\right]
=
\left[
\begin{array}{c}
1\\
0\\
0\\
1
\end{array}
\right]
\end{array}
$$
\begin{center}
\caption{Playing on the vertices $i_2$ and $i_3$ changes the state of the vertices $e_1$ and $e_4$}
\end{center}
\end{figure}
Theorem \ref{thm:car} shows that $A_D.X=0$ \emph{iff} $X=0$, so the rank of $A_D$ is $d$ (note that $d\le n-d$).
Thus the rank of its transpose $A_D^T$ is also $d$. Hence by pressing only on the vertices in $V\setminus D$, all the possible configurations for the vertices in $D$ can be reached.$\hfill \Box$
   
\end{proof}

This theorem shows that finding an optimal $d$-locally evenly seen set corresponds to finding a minimal set $D$ such that $A_D$  has 0 as eigenvalue, which implies the following corollary:
\begin{corollary} Given a graph $G=(V,E)$, 
$\delta_{loc}(G) \le \min(\lfloor |V|/2 \rfloor,\emph{rank}(A))$ where $A$ is the adjacency matrix of $G$.
\end{corollary}
Note that the relationship between the $A_X$ adjacency matrices and the local complementation has already bee pointed out in \cite{B87} where it was shown that the \emph{connectivity function} $c:X \rightarrow \mathrm{rank}(A_X)$ is invariant by local complementation.
\section{Preparation of Graph States}
 This section is dedicated to graph state preparation: for a given graph $G$, the preparation of $\ket{G}$ consists in applying some primitive operations, like unitary transformations or projective measurements, in order to obtain $\ket{G}$ starting from an initial state, which is an unknown state if  projective measurements are allowed and a known but not entangled state otherwise. The efficiency of a graph state preparation depends on the three following parameters:
 \begin{itemize}
 \item the number of ancillary qubits $\mathcal{N}$, 
 \item the size of the primitive operators $\mathcal{S}$,
 \item the logical depth $\mathcal{L}$ (i.e. the duration of the preparation). 
 \end{itemize}
 Since these parameters are not independent, there exist six different measures of complexity to compare two algorithms of preparation, according to the six lexicographic order over these three parameters. For instance if two algorithms of graph state preparation $\mathcal{A}_1$ and  $\mathcal{A}_2$ are compared according to the measure $1$-$\mathcal{S}, 2$-$\mathcal{L},3$-$\mathcal{N}$, first the size of the operators used by $\mathcal{A}_1$ and  $\mathcal{A}_2$ are compared, in case of equality, the logical depths are compared, and so on.
 
  The objective of this section is to present an exhaustive study of the efficiency of graph states preparation. For each complexity measure, an algorithm which optimizes the two most important parameters is presented. 
  
  Concerning the primitive operators, two cases are taken into account: the case where both unitary transformations and projective measurements are used and combined, and the case where only the projective measurements are allowed. This restriction to the projective measurements is motivated by one of the main applications of the graph state formalism which is measurement-based quantum computation and especially the One-Way Quantum Computer \cite{R03}. Notice that the restriction to projective measurements is particularly relevant when no ancillary qubits are used \ref{lem:pm}.

Two subroutines (\emph{edge-by-edge} and \emph{vertex-by-vertex}) for preparing a graph state, which are largely used in the algorithms of graph state preparation are presented. The first consists in adding to or removing  an edge from the associated graph of a given graph state and the second consists in adding a vertex and its neighborhood to the associated graph of a given graph state. The complexity of these methods is detailed below. 

\subsection{Edge by edge}

For a given graph state $\ket{G}$, where $G=(V,E)$, the application of the unitary transformation $C_Z$ (Controlled-$Z$) on some qubits $a$, $b$ leads to a graph state $\ket{G'}$, where $G'=(V,E\Delta (a,b))$. This operation consists in adding (removing) the edge $(a,b)$. This method is $LP$-robust, since if $C_Z$ is applied to some qubits $a$ and $b$ of a state $\ket{\phi}$ which is $LP$-equivalent to a graph state $\ket{G}$, where $G=(V,E)$, the resulting state $\ket{\phi'}$ is $LP$-equivalent to $\ket{G'}$ where $G'=(V,E\Delta (a,b))$.

This application of $C_Z$ can be simulated  with measurements only, up to known Pauli operators \cite{P04}. This simulation requires an additional qubit $c\notin V$, and leads to a state equal to $\ket{G'}$ up to a Pauli operator. This measurement based version of the  \emph{edge-by-edge} method is also $LP$-robust.

\subsection{Vertex by vertex}

For a given graph state $\ket{G}$, a sequence of measurements (see Lemma \ref{sm}) leads to a state which is $LP$-equivalent to $\ket{G'}$, where $G'$ is obtained from $G$ by adding a vertex $v$ and $k$ edges between $v$ and some vertices in $G$.

\begin{lemma}
\label{sm}
For any graph state $\ket{G}$ with $G=(V,E)$, a measurement in the computational basis on an additional qubit $v\notin V$ followed by a measurement on $(| N_G(v)| +1)$ qubits according to the observable $X^{(v)} \bigotimes_{k\in N_v} Z^{(k)}$, where $N_v \subset V$, leads to a state which is $LP$-equivalent to $\ket{G'}$, where $G'=(V\cup\{v\}, E\cup(\cup_{k\in N_v} (v,k)))$.

\end{lemma}

\begin{proof}

By defintion,$\ket{G}$ is stabilized by $\{g_j,j\in V\}$. After the $Z$-measurement of an additional qubit $v\notin V$, the state of the register $V\cup \{v\}$ is stabilized by  $\{ g'_j,j\in V \cup \{v\} \}$, with $g'_j= g_j$ if $j\in V$ and $g'_v=(-1)^{l}Z^{(v)}$, where $(-1)^{l}$ is the classical outcome of the measurement.

A stabilizer $\{ g''_j,j\in V \cup \{v\} \}$ of the state $\ket{\phi}$ after the $X^{(v)} \bigotimes_{b\in N_G(v)} Z^{(b)}$-measurement is obtained according to the evolution rules of stabilizers \cite{NC00}, where $ g''_j= g_j$ if $j\in V \setminus N_G(v)$, $g''_j=(-1)^lg_j.Z^{(v)}=(-1)^lX^{(j)}\bigotimes_{k\in N_G(j)\cup \{v\}} Z^{(k)}$ if $j\in N_G(v)$ and $g''_v=(-1)^{l'} X^{(v)} \bigotimes_{k\in N_G(v)} Z^{(k)}$, where $(-1)^{l'}$  is the classical outcome of the measurement.

Thus there exists $G'$ such that $\ket{\phi}\equiv_{LP}\ket{G'}$. Moreover $G'$ is obtained by adding to $G$ a vertex $v$ and $k$ edges between $v$ and each element of $N_v$.$\hfill \Box$
\end{proof}

Notice that this  \emph{vertex-by-vertex} method is also $LP$-robust.

\section{Optimization}

\subsection{Preparation using operators on two qubits only}
\label{section:SO}
Since local unitary transformations and local projective measurements do not create entanglement, operations on at least two qubits are needed. Furthermore, since unitary transformations on two qubits are universal for quantum computation \cite{NC00}, any graph state can be prepared using unitary transformations on two qubits only.

If only projective measurements are allowed, a similar result on the universality of projective measurements on two qubits can be found in \cite{P04}.

Thus in any case (\emph{i.e.} unitary transformations allowed or not), operations on two qubits are necessary and sufficient to prepare any graph state. Under the constraint of using operations on at most two qubits, we present first the optimization of the number of ancillary qubits, then the optimization of the logical depth.

\subsubsection{Optimization of the number of ancillary qubits using two-qubit operators}~

If unitary transformations are allowed, the \emph{edge-by-edge} method leads to an algorithm $\mathcal{A}_1$ using only the two-qubit unitary transformation $C_Z$ and no ancillary qubits. According to the complexity measure $1$-$\mathcal{S},2$-$\mathcal{N},3$-$\mathcal{L}$, $\mathcal{A}_1$ is optimal in terms of the size of the primitive operators and in terms of the number of ancillary qubits. In order to improve the logical depth of this algorithm, a maximal parallelization of the operations is done: for a given graph  $G$ with $n$ vertices, $G$ is edge-colored using $\chi'(G)$ colors where $\chi'(G)$ is the edge chromatic number of $G$. An no-edge graph state with $n$ vertices is prepared in one step, then for each color $c$, all $c$-colored edges are created in parallel by the \emph{edge-by-edge} method. This parallelization is allowed because the applied operations act on distinct qubits. The logical depth of this algorithm is $O(\chi'(G))$. These kinds of algorithms can be found for instance in \cite{L04}.

Since the preparation may be done up to a local unitary transformation, the algorithm can be improved by preparing not directly a graph $G$ but an equivalent graph $G'$ which has a minimal edge-chromatic number. For a given graph $G$, let $\chi'_{loc}(G)=min_{\ket{G'}\equiv_{LU}\ket{G}}(\chi'(G'))$. Thanks to Theorem \ref{vdn}, $\chi'_{loc}(G)$ is also the minimal edge-chromatic number under local complementation. Thus, instead of preparing directly the graph $G$, the graph $G'$ is prepared where $\ket{G'}$ and $\ket{G}$ are $LU$-equivalent and $\chi'(G')=\chi'_{loc}(G)$. In this case the logical depth of the preparation is $O(\chi'_{loc}(G))$.

Now, the case where unitary transformations are not allowed is taken into account: 

\begin{lemma}
If only $2$-qubit measurements are allowed, then the minimal number of ancillary qubits required for  graph state preparation is one. 
\end{lemma}
\begin{proof}
An algorithm using only one ancillary qubit  consists in using the measu\-rement-based version of the \emph{edge-by-edge} method.

As a consequence of Theorem \ref{thm:delt}, there exists some graph which cannot be prepared without ancillary qubit using only 2-qubit measurements. Thus, if only $2$-qubit measurements are allowed, then the minimal number of ancillary qubits required for  graph state preparation is one. $\hfill \Box$

\end{proof}
 
\begin{lemma}

Any graph state $\ket{G}$, where $G=(V,E)$, can be prepared with one ancillary qubit, using only $2$-qubit measurements with a  logical depth  $\frac{| E |}{\sqrt{| V| }}$. 

\end{lemma}  
  
\begin{proof}

 The algorithm of preparation is composed of two stages: in a first stage a set $K$ of $k$ qubits are used as ancillary qubits to add the edges which are not incident to $K$, then the vertices incident to $K$  are added using  the ancillary qubit $v$. Notice that a set of $m_i$ non-incident edges can be added using $k$ qubits as ancillary qubits in $\lfloor m_i/k \rfloor +1$ steps. Thus an edge-coloration in $\chi'$ colors of the edges non incident to $K$ leads to a preparation of the edges non-incident to $K$ in $ \sum_{i=1}^{\chi '} \lfloor m_i/k \rfloor +1\le \chi'+m(k+1)$ steps. 

The set $K$ is composed of the vertices which have a smallest degree, thus their average degree is less than $\frac{m}{n}$. The logical depth of the second stage is in this case upper bounded by $km/n$ (using the \emph{edge-by-edge} method). The optimal choice of the parameter $k$ is given by equalizing    $\chi'+m/k$ and $km/n$, and gives a logical depth $\frac{2m^2}{\chi'n+n\sqrt{4m^2/n+\chi'^2}}=O(\frac{m^2}{\sqrt{m^2n+\chi'^2n^2}})$ which can be upper bouded by $\frac{m}{\sqrt{n}}$.$\hfill \Box$
\end{proof} 

 Since the preparation of $\ket{G}$ may be done up to a local unitary transformation, instead of preparing the graph state $\ket{G}$, an equivalent graph state $\ket{\tilde{G}}=(\tilde{V},\tilde{E})$ is prepared  with an algorithm such that $$\frac{| \tilde{E} |}{\sqrt{| \tilde{V}| }}=min\{\frac{| E' |}{\sqrt{| V'| }}\, | \,  G'=(V',E') \, \mathrm{and} \, \ket{G'}\equiv_{LU}\ket{G}\}$$ 
\subsubsection{Optimization of the logical depth using two-qubit operators} ~

In this section the objective is to find an algorithm which is optimal according to the complexity measure $1$-$\mathcal{S},2$-$\mathcal{L},3$-$\mathcal{N}$. First, both unitary transformations and projective measurements are allowed.

\begin{figure}
\begin{center}
\includegraphics[width=10cm]{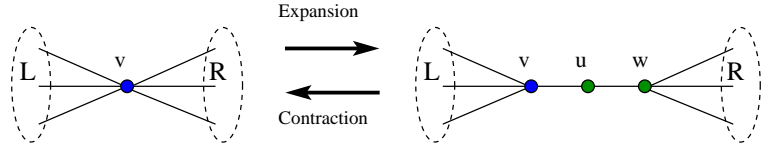}
\caption{Graph Expansion and Graph Contraction}
\label{contr}
\end{center}

\end{figure}

The algorithm presented in the previous section uses two-qubit unitary transformations  and has a linear logical depth in $\chi'(G)$, the edge-chromatic number 
of the graph to prepare. 
To optimize the duration of the preparation, the following graph transformations are used (see Figure \ref{contr}):
\begin{itemize}
\item \emph{Graph expansion:} Given a graph $G=(V,E)$, the graph expansion of $G$ with respect to a vertex $v$ and a subset $R \subseteq N_G(v)$ of neighbors of $v$, is the graph $G'=(V\bigcup \{u,w\},E')$ where $E'=(E\setminus \{(v,v_i), v_i \in R\} \bigcup \{(v,u),(u,w)\}\bigcup$  $\{(w,v_i), v_i \in R\}$.
\item The reverse transformation is a \emph{graph contraction}.
\end{itemize} 

\emph{Graph expansion} permits to decrease the degree, indeed:
\begin{property}
For any graph $G$, there exists a sequence of expansions which leads to a graph $G'$ such that $\Delta(G')\le 3$.
\end{property}

\begin{proof}
A constructive proof consists in expanding recursively all the vertices with a degree greater than $4$. $\hfill \Box$ 
\end{proof}

\begin{figure}

\begin{center}
\includegraphics[width=10cm]{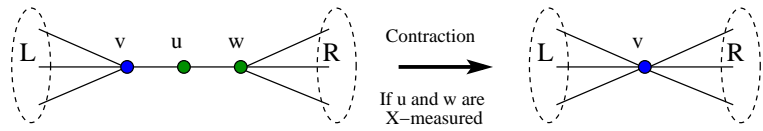}
\caption{Graph State Contraction}
\label{gsc}
\end{center}
\end{figure}

These graph transformations are particularly relevant in graph state preparation because a graph contraction can be realized on a graph state by means of one-qubit measurements only, see Figure \ref{gsc}. This graph state contraction is $LP$-robust.

 From this property comes the following  algorithm $\mathcal{A}_2$ of graph state preparation.  Given a graph $G$,
 \begin{itemize}
 \item $G$ is expanded into a graph $G'$ such that $\Delta(G')\le 3$, notice that this stage is entirely classical.
 \item $\ket{G'}$ is prepared according to the previous algorithm using two-qubit unitary transformations only and a logical depth linear in $\chi'(G')$ the edge-chromatic number of $G'$. Since for any graph $G$, $\chi' (G)\le \Delta(G)+1$, this stage needs a constant logical depth.  
 \item All the ancillary qubits, corresponding to the vertices added during the expansion, are measured according to the observable $X$ (i.e. in the basis $\{\ket{+},\ket{-}\}$) at the same time. The logical depth of this stage is $1$ and it leads to a state $LU$-equivalent to  the graph state $\ket{G}$.
 \end{itemize}
 
 $\mathcal{A}_2$ permits to prepare any graph state using $2$-qubit primitive operators with a constant logical depth. Thus the first two parameters of the complexity measure  $1$-$\mathcal{S},2$-$\mathcal{L},3$-$\mathcal{N}$ are optimized. An upper bound of the complexity in terms of ancillary qubits is given by the following lemma:

\begin{lemma}
\label{lemma:A2}
Any graph state $\ket{G}$ can be prepared in a constant time using only two-qubit operators. The number of needed ancillary qubits is upper-bounded by $min_{\ket{G'}\equiv_{LU}\ket{G}}(4|E'|-2| V'|),\, where\ G'=(V',E')$.
\end{lemma}

\begin{proof}
In order to prepare $\ket{G}$, given a graph $G=(V,E)$ with the algorithm $\mathcal{A}_2$, each graph state contraction costs two ancillary qubits to lower the degree by one. The number of qubits required by this algorithm is then:
$\sum_{d_i>3} 2(d_i-3)$ where $d_i$ is the degree of the $i^{th}$ vertex.
This can be upper-bounded by 
$$\sum_{d_i>3} 2(d_i-3)+2 \sum_{d_i \le 3}d_i - 2 \sum_{d_i \le 3}d_i\le 2 \big(\sum d_i\big) - 2| V|  =
 4| E| -2| V |$$ 
 
 Since the preparation of $\ket{G}$ may be done up to a local unitary transformation, instead of preparing the graph state $\ket{G}$, an equivalent graph state $\ket{\tilde{G}}=(\tilde{V},\tilde{E})$ is prepared  with the algorithm $\mathcal{A}_2$ such that $$4|\tilde{E}|-2|\tilde{V}|=min \{(4| E'|-2| V'|)\, | \,  G'=(V',E') \, \mathrm{and} \, \ket{G'}\equiv_{LU}\ket{G}\}$$ $\hfill \Box$
 
 \end{proof}

Consider now the case where unitary transformations are not allowed:

\begin{lemma}
\label{lemma:A2meas}
If only $2$-qubit measurements are allowed, then any graph state $\ket{G}$ can be prepared in a constant time.  
\end{lemma}

\begin{proof}
The proof is based on the simulation of each unitary transformation of the algorithm $\mathcal{A}_2$ by means of projective measurements. The number of ancillary qubits can be upper-bounded using a method similar to  \emph{Lemma \ref{lemma:A2}}.
\end{proof}

\subsection{Preparation in a constant time}

\emph{Lemma \ref{lemma:A2}} proves that any graph state can be prepared in a constant time. 

 Under this constraint, first the optimization of the size of the primitive operators, then the optimization of the number of ancillary qubits are optimised:

 \subsubsection{Optimization of the size of the primitive operators for constant time preparations}~

The objective is to find an algorithm which is optimal according to the complexity measure $1$-$\mathcal{L},2$-$\mathcal{S},3$-$\mathcal{N}$. 

If both unitary transformations and projective measurements are allowed, \emph{Lemma \ref{lemma:A2}} proves that any graph state can be prepared in a constant time using only two-qubit operators. Thus the first two parameters of the complexity measure $1$-$\mathcal{L},2$-$\mathcal{S},3$-$\mathcal{N}$ are optimized.

If unitary transformations are not allowed, \emph{Lemma \ref{lemma:A2meas}} proves that any graph state can be prepared in a constant time using two-qubit measurements only.

 \subsubsection{Optimization of the number of ancillary qubits for constant time preparations}~

The objective is to find an algorithm which is optimal according to the  complexity measure $1$-$\mathcal{L},2$-$\mathcal{N},3$-$\mathcal{S}$. 

For any graph $G=(V,E)$, there exists a unitary transformation $U$ which transforms the state $\ket{+}^{\otimes | V|}$ into $\ket{G}$. $U$ is nothing but the composition of the unitary transformations used in the \emph{edge-by-edge} method. Such a constant time preparation needs no ancillary qubits, so the first two parameters of the complexity measure  $1$-$\mathcal{L},2$-$\mathcal{N},3$-$\mathcal{S}$ are optimized. This preparation requires a $| V |$-qubit unitary transformation.

If unitary transformations are not allowed, a similar preparation can be done, where the unitary transformation $U$ is replaced by a projective measurement in the basis composed of the states $\sigma \ket{G}$, for all $|V|$-qubit Pauli operators $\sigma$.

 \subsection{Preparation without Ancillary Qubits}

 Since, in section \ref{section:SO}, an algorithm based on the \emph{edge-by-edge} method, permits to prepare any graph state without ancillary qubits,  the minimal number of ancillary qubits required to prepare any graph state is zero.  
 Under the constraint of preparing any graph state without ancillary qubits, first the optimization of the logical depth, then the optimization of the size of the primitive operators are presented.  
 
  \subsubsection{ Optimization of the logical depth without ancilla qubit}~
  
  The objective is to find an algorithm which is optimal according to the complexity measure $1$-$\mathcal{N},2$-$\mathcal{L},3$-$\mathcal{S}$.
  
  The algorithm used to optimize the first two parameters of the complexity measure $1$-$\mathcal{L},2$-$\mathcal{N},3$-$\mathcal{S}$, optimizes also the first two parameters of the measure 
 $1$-$\mathcal{N},2$-$\mathcal{L},3$-$\mathcal{S}$: a constant time preparation without ancillary qubit.

  \subsubsection{ Optimization of the size of the primitive operators without ancilla qubit}~
  
The objective is to find an algorithm which is optimal according to the complexity measure $1$-$\mathcal{N},2$-$\mathcal{S},3$-$\mathcal{L}$. 

   If both unitary transformations and projective measurements are allowed, the algorithm used to optimize the first two parameters of the complexity measure $1$-$\mathcal{S},2$-$\mathcal{N},3$-$\mathcal{L}$, optimizes also the first two parameters of the measure 
 $1$-$\mathcal{N},2$-$\mathcal{S},3$-$\mathcal{L}$: a preparation without ancillary qubits and using two-qubit operations only. 
 
 The case where unitary transformations are not allowed is particularly relevant. An algorithm which permits to prepare any graph state $\ket{G}$ without ancillary qubits consists in considering a vertex $v\in V$ such that the degree of $v$ is minimal, i.e. $\delta(v)=\delta(G)$, then
 \begin{itemize}
 \item each edge which is not incident to $v$ is added using the measurement-based version of the \emph{edge-by-edge} method, where $v$ is used as ancillary qubit, and 
 \item the vertex $v$ is added using the \emph{vertex-by-vertex} method, this method requires a projective measurement on $\delta(G)+1$ qubits.
 \end{itemize} 
 
 Since the preparation may be done up to a local unitary transformation, the algorithm can be improved by preparing not directly a graph state $\ket{G}$ but an equivalent graph state $\ket{G'}$, where $G'$ has a smaller  minimal degree than  $G$. 
Thus, instead of preparing directly the graph state  $\ket{G}$, the graph state $\ket{G'}$ is prepared where $\ket{G'}$ and $\ket{G}$ are $LU$-equivalent and $\delta(G')=\delta_{loc}(G)$. In this case the size of the primitive operators is $\delta_{loc}(G)+1$.
 
 The following theorem proves that this algorithm optimizes the first two parameters of the complexity measure $1$-$\mathcal{N},2$-$\mathcal{S},3$-$\mathcal{L}$:
 
\begin{theorem}
\label{thm:delt}
A measurement-based preparation of any graph state $\ket{G}$ without ancillary qubit needs measurements on at least $\delta_{loc}(G)+1$ qubits.
\end{theorem}
 
\begin{proof}
Let $G_0=(V_0,E_0)$ be a graph such that $\delta(G_0)=\delta_{loc}(G)$ and $\ket{G_0}$ LU-equivalent to $\ket{G}$.
We assume that there exists a preparation of $\ket{G_0}$ up to $LU$-equivalence, using only measurements on at most $\delta(G_0)$ qubits, and we consider the last measurement of this preparation. The rest of the proof consists in showing that this last measurement $\mathcal{O}$ is nothing but $identity$, which implies that such a preparation using measurements on at most $\delta(G_0)$ qubits does not exist. 

Without loss of generality we assume that the classical outcome of $\mathcal{O}$ is $1$, i.e. $\mathcal{O}\ket{\phi}=\ket{\phi}$, where $\ket{\phi}$ is the state produced by the preparation.

We first assume that only Pauli measurements are used during the preparation, the general case is discussed after.

 Since only Pauli measurements are used, the preparation produces a \emph{stabilizer} state $\ket{\phi}$.
 Since $\ket{\phi}$ and $\ket{G_0}$ are both stabilizer states and are LU-equivalent, $\ket{\phi}$ and $\ket{G_0}$ are LC-equivalent, \emph{i.e.} there exists a local Clifford transformation $C$ such that $\ket{\phi}=C\ket{G_0}$ (one can show that the hypothesis of the theorem proposed by  Van den Nest in \cite{VDM04} are verified). 

$$\mathcal{O}\ket{\phi}=\ket{\phi} \iff  \mathcal{O}C\ket{G_0}=C\ket{G_0} \iff  C^{\dagger}\mathcal{O}C\ket{G_0}=\ket{G_0}$$

Since $C$ is a local Clifford transformation, $W=C^{\dagger}\mathcal{O}C$ is a Pauli transformation on $\delta(G_0)$ qubits such that $\ket{G_0}$ is a fixpoint of $W$. Thanks to Theorem \ref{thf}, $G_0$ has no $\delta(G_0)$-locally evenly seen set. So the set $D$ composed of the $\delta(G_0)$ qubits measured during the last measurement is seen by a qubit $v_0 \in V_0\backslash D$ such that $| D_{v_0}|$ is odd where $D_{v_0}=N_G(v_0)\cap D$.
 
 Since $W\ket{G_0}=\ket{G_0} \iff Wg_{v_0}\ket{G_0}=g_{v_0}\ket{G_0}\iff Z^{(D_{v_0})}WZ^{(D_{v_0})}\ket{G_0}=\ket{G_0}$ 
and since $W$ is a Pauli operator, $W$ and $Z^{(D_{v_0})}$ commute. 
So there exists $v_1\in D_{v_0}$ such that $W^{(v_1)}\in \{I,Z\}$.
 
 Thus $W\ket{G_0}=\ket{G_0} \iff Wg_{v_1}\ket{G_0}=g_{v_1}\ket{G_0}\iff X^{(v_1)}WX^{(v_1)}\ket{G_0}=\ket{G_0}$ so $W$ and $X{(v_1)}$ commute, which implies that $W^{(v_1)}=I$.
 Applying Theorem \ref{thf} by induction on
 $D\backslash \{v_1\}$, one can prove that $\forall v \in D, W^{(v)}=I$, so $W=I$.

Now we consider the general case where not only Pauli measurements are used. Since the prepared state $\ket{\phi}$ is LU-equivalent to $\ket{G_0}$, there exists a local unitary transformation $U$ such that $\ket{\phi}=U\ket{G_0}$. $W=U^{\dagger}\mathcal{O}U$ is a transformation on $\delta(G_0)$ qubits such that $\ket{G_0}$ is a fixpoint of $W$. The application of a unitary transformation $X$ on a qubit $v\in V_0\backslash D$ changes the sign of  all the operators $g_{v'}$ of the stabilizer of $\ket{G_0}$, for all $v'\in N_G(v)$. That is why the application of the unitary transformation $X$ can be interpreted in terms of a combinatorial game on graphs called \emph{$\sigma$-game} already introduced in section \ref{grph}: 

All the vertices are labeled by an assignment function $\mu : V_0\to\{0,1\}$. With each assignment function $\mu$ is associated the quantum state stabilized by $S_{\mu}=\{(-1)^{\mu(v)}g_v,v\in V_0\}$, where $\{ g_v,v\in V_0\}$ is the stabilizer of $\ket{G_0}$. One can verify that this correspondence between pressing a vertex $v$ in the $\sigma$-game and applying $X$ on the qubit $v$ of the graph state is valid, i.e. if pressing $v$ leads to a configuration $\mu$, then the application of $X$ on $v$ leads to a state stabilized by $S_{\mu}$.

Let $D$ be the set of the $\delta(G_0)$ measured qubits. Thanks to Theorem \ref{thf2}, for any configuration on $D$ (i.e. for any $\mu : D\to\{0,1\}$) there exists a subset $R_{\mu}\subset (V_0\backslash D)$ such that if $X$ is applied on all the qubits of $R_{\mu}$, the obtained state is stabilized by $\tilde{S}_{\mu}=\{(-1)^{\mu(v)}g_v,v\in D\}\cup \{ g_v,v\in V_0\backslash D\}$. Since $D$ and $R_{\mu}$ are distinct sets, one can easily prove that the state stabilized by $\tilde{S}_{\mu}$ is a fixpoint of $W$.

Then we assume that all the qubits of $V_0\backslash D$ are measured according to $Z$, and that the classical outcome of each measurement is $1$. This sequence of measurements leads to a state on $\delta(G_0)$ qubits stabilized by $S'_{\mu}=\{(-1)^{\mu(v)}g_v^{(D)},v\in D\}$, where $g_v^{(D)}$ is the operator $g_v$ restricted to the qubits of $D$. Since the qubits of $D$ are not measured, one can prove that the state stabilized by $S'_{\mu}$ is a fixpoint of $W$.

\begin{claim}
 For any $\mu, \mu'$, if $\mu\neq \mu'$ then the states $\ket{\phi_1}$ and $\ket{\phi_2}$ respectively stabilized by $S'_{\mu}$ and $S'_{\mu'}$ are orthogonal.
\end{claim}
\begin{proof}
If $\mu\neq \mu'$, there exists $v\in D$ such that $\mu(v)=-\mu'(v)$, $\langle \phi_1\ket{\phi_2}=(\langle \phi_1| \mu(v))(-\mu(v)\ket{\phi_2})=-\langle \phi_1\ket{\phi_2}$, which implies that $\langle \phi_1\ket{\phi_2}=0$.

\end{proof}

Since there exist $2^{| D|}$ different configurations $\mu:D \to \{0,1\}$, the corresponding set of qubits composed of $2^{| D|}$ mutually orthogonal states is a basis of the subspace composed of $| D| $ qubits. Since for any state $\ket{\phi}$ of this basis $W\ket{\phi}=\ket{\phi}$, this means that $W$ is nothing but identity.$\hfill \Box$

\end{proof}

The previous algorithm is optimal in terms of ancillary qubits and in terms of the size of the measurements used. In order to improve the logical depth of this algorithm a parallelization of the first stage, which consists in adding some vertices using the measurement based version of the \emph{edge-by-edge} method, can be done:

\begin{lemma}
\label{lem:pm}
Any graph state $\ket{G}$ can be prepared without ancillary qubits, using only projective measurements on $\delta_{loc}(G)+1$ qubits. The logical depth of this preparation can be upper bounded by $\frac{| E' |}{\sqrt{| V'| }}$, where $G'=(V',E')$ is a graph such that $\ket{G'}\equiv_{LU}\ket{G}$ and $\delta(G')= \delta_{loc}(G)$.

\end{lemma}  
  
\begin{proof}
Let $G'$ be a graph such that $\ket{G'}\equiv_{LU}\ket{G}$ and $\delta(G')= \delta_{loc}(G)$. Let $m=| E' |$ and $n=| V' | $, let $v\in V'$ be a vertex such that $\delta(v)=\delta(G')$. The algorithm of preparation is composed of two stages: a first stage consists in adding each edge which is not incident to $v$ using the measurement-based version of the \emph{edge-by-edge} method, the second stage consists in adding $v$ using the \emph{vertex-by-vertex} method. 

The first stage is separated into two parts: a set $K$ of $k$ qubits are used as ancillary qubits to add the edges which are not incident to $K$, then the vertices incident to $K$ but not to $v$ are added using $v$ as ancillary qubit. Notice that a set of $m_i$ non-incident edges can be added using $k$ qubits as ancillary qubits in $\lfloor m_i/k \rfloor +1$ steps. Thus an edge-coloration in $\chi'$ colors of the edges non incident to $K$ leads to a preparation of the edges non-incident to $K$ in $ \sum_{i=1}^{\chi '} \lfloor m_i/k \rfloor +1\le \chi'+m/k$ steps. 

The set $K$ is composed of the vertices which have a smallest degree, thus their average degree is less than $\frac{m}{n}$. The logical depth of the second stage is in this case upper bounded by $km/n$. The optimal choice of the parameter $k$ is given by equalizing   $\chi'+m/k$ and $km/n$, and gives as logical depth $\frac{2m^2}{\chi'n+n\sqrt{4m^2/n+\chi'^2}}=O(\frac{m^2}{\sqrt{m^2n+\chi'^2n^2}})$ which can be upper bouded by $\frac{m}{\sqrt{n}}$. So the logical depth of the first stage is upper bounded by $\frac{m}{\sqrt{n}}$.

The second stage is realized in a single step, thus the logical depth of the preparation is upper bounded by $\frac{m}{\sqrt{n}}$.
$\hfill \Box$
\end{proof}
\section*{Conclusion}

For each complexity measure of graph state preparation, which depends on three parameters (the size of the operators, the number of ancillary qubits, and the logical depth), an algorithm optimizing the first two parameters is presented in this paper. In particular, any graph state can be prepared in a constant time, using operators on two qubits only. Moreover we prove that a measurement-based preparation of any graph state $\ket{G}$ without ancillary qubits, requires measurements on  $\delta_{loc}(G)+1$ qubits, where $\delta_{loc}(G)$ is the minimal degree of $G$ under local complementation.

To complete the characterization of the complexity of graph state preparation, it would be interesing to find lower bounds for the third parameter. It is also important to consider
the classical time for minimizing some graph parameters under local complementation (the minimal degree, the number of edges, the maximal degree, \ldots ). For the minimal degree, using the results of section \ref{grph}, a suggested method to tackle the problem would be to find the complexity of computing a minimal $d$-locally evenly seen set.

\end{document}